\documentclass[examplefnt,biber]{nowfnt} 
\usepackage{
geometry}
\usepackage{setspace}
\usepackage[utf8]{inputenc}
\usepackage{url}

\usepackage{amssymb}
\title{Decentralized Finance: Protocols, Risks, and Governance}

\subtitle{Decentralized Finance}

\maintitleauthorlist{
Agostino Capponi  \\
Department of Industrial Engineering and Operations Research, \\ Columbia University, \\ %, 500 West 120th Street, New York, NY 10027 \\
ac3827@columbia.edu
\and
Garud Iyengar \\
Department of Industrial Engineering and Operations Research, \\ Columbia University,\\ %500 West 120th Street, New York, NY 10027 \\
garud@ieor.columbia.edu
\and
Jay Sethuraman \\
Department of Industrial Engineering and Operations Research\\, Columbia University,\\ %500 West 120th Street, New York, NY 10027 \\
jay@ieor.columbia.edu
}

\usepackage{mwe}

\author[1]{Capponi,Agostino}
\author[2]{Iyengar,Garud}
\author[3]{Sethuraman,Jay}

\affil[1]{Department of Industrial Engineering and Operations Research, Columbia University; %, 500 West 120th Street, New York, NY 10027;
ac3827@columbia.edu}
\affil[2]{Department of Industrial Engineering and Operations Research, Columbia University; %, 500 West 120th Street, New York, NY 10027; 
garud@ieor.columbia.edu}
\affil[3]{Department of Industrial Engineering and Operations Research, Columbia University; %, 500 West 120th Street, New York, NY 10027; 
jay@ieor.columbia.edu}

\articledatabox{\nowfntstandardcitation}

\begin{document}
\makeabstracttitle

\begin{abstract}
Financial markets are undergoing an unprecedented transformation. Technological advances have brought major improvements to the operations of financial services. 
 While these advances promote improved accessibility and convenience, traditional finance shortcomings like lack of transparency and moral hazard frictions  continue to plague centralized platforms, imposing societal costs.

 In this paper, we argue how these shortcomings and frictions are being mitigated by the decentralized finance (DeFi) ecosystem. We delve into the workings of smart contracts, the backbone of DeFi transactions, with an emphasis on those underpinning token exchange and lending services.

We highlight the pros and cons of the novel form of decentralized governance introduced via the ownership of governance tokens. Despite its potential, the current DeFi infrastructure introduces operational risks to users, which we segment into five primary categories: consensus mechanisms, protocol, oracle, frontrunning, and systemic risks.

We conclude by emphasizing the need for future research to focus on the scalability of existing blockchains, the improved design and interoperability of DeFi protocols, and the rigorous auditing of smart contracts.

\end{abstract}

\chapter{Introduction}

%These platforms are profit maximizing entities, which extract make profits by charging fees to users. 

%Fees are in different forms (spreads, or fees) The financial industry has experienced a first wave of disruption over the last two decades through the increasing penetration of technology and digitalization into existing platforms and services. A second wave of disruption is ongoing and is been triggered by the development of distributed ledger technologies. Those technologies are facilitating the transition of financial services from a centralized to a decentralized structure. Decentralized Finance (DeFi) has the potential to overcome inefficiencies and limitations of centralized finance, and further improve the provision of services.

%. These fees vary based on the products, ranging from account fees (monthly maintenance charges, minimum balance fees, overdraft fees, and non-sufficient funds [NSF] charges), safe deposit box fees, and late fees.

Financial services have traditionally been provided through centralized
platforms. Notable instances include Visa and Mastercard, the world's leading payment processing networks; Nasdaq, the globe's premier stock exchange; Vanguard and Blackrock, renowned for their investment and brokerage services; and JP Morgan Chase, offering a spectrum of banking services.  While the centralized financial system is essential to provide intermediation services to the real economy, it can also be exclusionary and impose hefty costs on users. Take credit card companies as an example: they impose processing fees ranging from 2\% to 4.35\% of the transaction's value. Moreover, commercial banks frequently apply considerable service charges and loan interest rates, yet offer low interest rates on customer deposits. The cost of financing can be prohibitive for small borrowers who may find loans or mortgages inaccessible, leading them to depend on credit cards that demand high interest rates. Even access to basic financial services, such as payment services, can be uncertain, particularly in less developed regions where a significant portion of the population remains unbanked.

The integration of technology into finance has sparked the rise of alternative service providers that alleviate some of these concerns. Emergent payment systems, like Square and Venmo, utilize mobile devices and internet technology to provide affordable, user-friendly payment solutions. These services are accessible to individuals and small businesses alike, democratizing access to financial systems

Peer-to-peer lending platforms, like Prosper and LendingClub, harness the power of technology to develop innovative lending marketplaces. These platforms offer an alternative to traditional credit sources by facilitating loans, primarily funded by institutional investors, to borrowers. Fintech brokerage platforms, such as Robinhood, enable users to execute commission-free trades of stocks, exchange-traded funds, and cryptocurrencies via mobile applications.

While these financial technology advancements offer increased convenience, inclusivity, and cost reductions, they also inherit several drawbacks associated with traditional centralized finance (CeFi). Firstly, decreases in transaction costs and enhancements in execution speed are not necessarily a given, as these fintech applications are constructed on the existing financial infrastructure. For instance, when receiving a payment through Venmo followed by a deposit into your bank account, a bank transfer must be initiated. Such transfers may take up to three business days to complete. Even though an instant transfer option exists, it remains a costly alternative, imposing a 1.75\% fee.

Most extant fintech platforms still function as centralized, profit-driven entities, which often leads to moral hazard issues and societal costs. Several of these platforms have faced criticisms for prioritizing their own profitability over customer interests. A notable example is Robinhood Markets, a firm offering commission-free investment services, which was found to have routed customer order flows to high-frequency trading firms rather than stock exchanges, as indicated in \citet{Robinhood2021}.

Similarly, the lack of transparency in some peer-to-peer (P2P) lending platforms' credit assessment methodologies can lead to defaults and investor losses. As profit-oriented entities, P2P companies primarily aim to enhance their profits. Consequently, despite lenders' desire to steer clear of high-risk borrowers, P2P platforms may entice borrowers to take larger loans by offering appealing interest rates and neglecting credit risk. This moral hazard dilemma contributed to the downfall of numerous P2P lending firms, particularly in China, as reported in \citet{p2pFintech}.

The advent of distributed ledger technologies presents an opportunity to alleviate some of the issues raised by centralized financial platforms, regardless of their integration of fintech enhancements. These technologies have the potential to further disrupt the financial service industry by facilitating the transition to a decentralized trading environment, also referred to as {\it decentralized finance} (DeFi). DeFi enables the provision of services such as exchanges, lending, derivatives trading, and insurance without the need for a centralized intermediary. The rest of the paper is organized as follows. Chapter~\ref{sec:DeFi} provides an overview of the DeFi ecosystem, with a focus on  exchanges, lending protocols, and the decentralized governance structure in place. Chapter~\ref{sec:operationalrisk} discusses the operational risks inherent in the design of smart contracts and the DeFi ecosystem. %Chapter~\ref{sec:tokenization} discusses the process of asset tokenization and the resulting benefits from an investor's perspective. 
We provide concluding remarks and directions for future research in Chapter~\ref{sec:conclusion}.
%Decentralized Finance (\ref{sec:conclsuions}.DeFi) has the potential to overcome inefficiencies and limitations of centralized finance, and further improve the provision of services.

%WHY DEFI HAS THE POTENTIAL TO INCREASE DEMOCRATIZATION, ETC>>>

\chapter{The Decentralized Finance (DeFi) Ecosystem}\label{sec:DeFi}
This chapter provides an introduction to the DeFi ecosystem and the underlying technological infrastructure that supports it. In Section~\ref{sec:definfr}, we 
discuss the blockchain infrastructure and introduce the smart contract functionality. In Section~\ref{sec:stablecoins}, we provide a categorization of stablecoins, one of the key pillars of the DeFi ecosystem. In Section~\ref{sec:dexes}, we examine the mechanics of decentralized exchanges. In Section~\ref{sec:lending}, we analyze decentralized borrowing and lending services. In Section~\ref{sec:decgov}, we discuss decentralized governance, and how it is implemented in the DeFi ecosystem. Throughout the chapter, we draw comparisons between DeFi services and traditional intermediaries, as well as with centralized forms of governance in the legacy financial system.

\section{Blockchain, Smart Contracts, and DeFi}\label{sec:definfr}

%DESCRIBE WHAT IT MEANS a peer-to-peer, decentralized, anonymous, and transparent network. AND MENTION THAT PROOF OF WORK AND PROOF CHAIN ARE WAYS TO IMPLEMENT BLOCKCHAIN
%ADD THE EVM AS WHAT IT ALLOWS TO IMPLEMENT SMART CONTRACTS

The first study of blockchain dates back to the work of \citet{Haber}. However, it has only been a few years later that blockchain became very popular thanks to \citet{bitcoin}. In his white paper, Nakamoto envisioned blockchain as the decentralized digital ledger to support the Bitcoin cryptocurrency. This ledger publicly records all executed transactions, with no central authority supervising the integrity of such transactions. The system relies on network of computer nodes to verify, update and store transactions in blocks, all linked to each other via a chain. This decentralized mechanism requires nodes to reach a consensus on the integrity of transactions and on the correct chain to which  the next block is appended. The first protocol proposed to achieve consensus was proof of work (PoW),  popularized by \citet{bitcoin}. This protocol incentivizes nodes to participate in the consensus task through a process called mining. Nodes of the network, also referred to as miners, compete to solve a computationally costly problem and the winner is  rewarded with newly minted coins plus fees paid by transactions appended in the block. %PoS replaces the PoW competition by randomly selecting stakeholders which validate and append the next block to the chain. 

The original vision of Nakamoto was for blockchain  to support peer-to-peer  payment transactions, essentially to enable users to transfer Bitcoin funds from their own digital addresses to other digital addresses. It soon became clear that the distributed ledger infrastructure could evolve to support a wider range of economic and financial applications, beyond payments and transfers. The breakthrough which made this possible was made by Ethereum (see \citet{Buterin}), a decentralized platform with its own blockchain whose native 
 cryptocurrency Ether is the second largest cryptocurrency by market capitalization after Bitcoin. Ethereum introduced two main technical innovations: (i) smart contracts, and (ii) the Ethereum virtual machine (EVM). 

 Smart contracts are digital contracts which are self-enforcing through automated execution, and whose terms are contingent on the reach of decentralized consensus. They facilitate the implementation of services and exchange of tokens in an algorithmically automated and conflict-free way. Moreover, leveraging the decentralized blockchain's power they guarantee enforceability without any need of a third party. For example, they can be programmed to guarantee locked in collateral for withdrawal of funds, or to make a payment to a digital address only if certain contingencies are met. %At present, smart contracts support a broad ecosystem of services, including lending, exchange, derivatives trading, and insurance services. 
 We refer to \citet{smartcontr} and \citet{HeCongRFS} for additional details on smart contracts.

To make smart contracts readable by the blockchain network, Ethereum created the so-called virtual machine (EVM). The EVM compiles smart contract code into a standard format known as ‘bytecode’, which is readable by the Ethereum network. As a result, transactions can be recorded by the nodes on the Ethereum network. Rather than as a distributed record-keeping ledger, Ethereum functions as a distributed state machine. The Ethereum's state is a large data structure which, in addition to holding all users' accounts and balances, also includes a machine state. The state changes from block to block according to prespecified rules, and the machine can execute arbitrary code.\footnote{For additional details on the Ethereum virtual machine, we refer to \url{https://ethereum.org/en/developers/docs/evm/}.}

%After Bitcoin’s success, it was time for the next generation of blockchain, which was brought about by Ethereum. Second-generation blockchain technology does more than just document transactions. Using self-executing agreements between two parties, called smart contracts, transactions are faster and more secure than first-generation blockchain technology. Another advantage of second-generation blockchain technology is that it acts more like a digital ecosystem instead of a system solely for transactions. Ethereum, for example, is a network that DeFi applications, games, and NFTs can run on. But with the advent of “The Merge”, Ethereum is shifting towards the next generation of blockchain: the third generation!

The EVM gave rise to the so-called second-generation blockchains that support decentralized finance (DeFi). Examples of DeFi services include decentralized exchanges based on automated market making, as well as decentralized lending and borrowing protocols such as Aave. As stated in \citet{CongInclusion}, the transaction volume in DeFi has increased enormously, from accounting less than 10\% of the transaction volume in 2017 to accounting for about 90\% since 2020. DeFi enable users to interact with smart contracts deployed on the blockchain. Users can access DeFi applications through  tokens. A token is a digital representation of an asset and is built on a specific blockchain. Tokens are typically distributed through an initial coin offering, i.e., a crowd sale whose objective is to raise funds.

Despite its popularity, there is at present not a formal definition of DeFi. \citet{GervaisCeFiDeFi} aims to establish such a definition by proposing a set of criteria to differentiate between CeFi and DeFi services (refer to Figure 1 in their paper). According to their classification, for a service to be considered DeFi, users must have full control over their assets, including custody and direct transaction capabilities without relying on any financial intermediary. Additionally, DeFi protocols should not allow any party to unilaterally censor transaction executions, and there should be no entity with the authority to unilaterally halt or censor the protocol's execution.

%At present, smart contracts support a broad ecosystem of services, including lending, exchange, derivatives trading, and insurance services.

\section{Stablecoins}\label{sec:stablecoins}
Stablecoins are digital assets designed to maintain a steady value in relation to a specified reference asset (see \cite{TreasuryWG}). They play an important role in the DeFi ecosystem, facilitating fund transfers across platforms and between users.

The term ``stablecoin'' groups together a diverse array of financial instruments, each employing distinct arrangements and collateral requirements to maintain the peg relation with respect to the underlying assets. Smart contracts are typically used to manage the creation and redemption of stablecoins. 

The most prevalent category within the stablecoin ecosystem comprises of those that are {\it collateralized by fiat currencies}. These stable coins allow DeFi market participants to avoid converting to and from fiat money at every turn. 
The two largest stablecoins by market capitalization are USDT and USDC, issued by Tether and Circle, respectively, which enable users to convert their traditional currency, such as USD, into an equivalent amount of USDT or USDC at a 1:1 ratio. The fiat currency received is held in reserve by the stablecoin issuer, providing the promise that users can later redeem their tokens for an equal value in fiat currency. There have been concerns regarding the stability of USDT's reserve assets. Tether has faced accusations of using unbacked reserves to support the value of its stablecoin, raising questions about its true collateralization. On the other hand, USDC is widely regarded as more stable and transparent. It is fully backed by collateral, and USDC tokens can be redeemed for USD at a 1:1 ratio through Circle's partnered regulated financial institutions. To ensure reserve adequacy, these institutions undergo monthly attestations and audits, providing greater confidence in the legitimacy of the reserves backing USDC.

The rapid expansion of fiat-backed stablecoins has raised financial stability concerns, because of the spillover effects between the crypto and the traditional financial systems. An illustrative instance occurred with Circle's USDC, where its price sharply dropped by over 15\% in March 2023 within a few hours, coinciding with the collapse of Silicon Valley Bank (for an in-depth analysis of stablecoin runs, refer to \cite{Mastablecoins}). Such events underscore the vulnerability of fiat-backed stablecoins to sudden market shocks. In the event of a run on these stablecoins, not only would stablecoin investors suffer direct losses, but it could also place significant strain on crucial asset markets such as deposit funding, Treasuries, and corporate bonds. 
The repercussions of such instability and potential runs on stablecoins have sparked widespread discussions about the necessity for comprehensive regulation (see, for instance,  \citet{IMFstablecoins} and \citet{ECBstablecoins}). 

A second class of stablecoins consists of {\it crypto-collateralized} stablecoins. Unlike fiat-backed stablecoins, these are backed by a reserve of other cryptocurrencies rather than fiat currencies. Collateral is locked in smart contracts rather than with financial institutions. Stabilizing algorithms are coded into the smart contracts to adjust or incentivize the supply and demand of stablecoins, thereby maintaining their pegs. Notable examples of crypto-backed stablecoins include DAI by MakerDAO, FRAX by the Frax Protocol, FEI by the FEI protocol (which closed at the end of 2022), and MIM by the Abracadabra.money platform.

A third class of stable coins are {\it algorithmic} stablecoins, which rely on algorithms to maintain their peg. A prominent example of such a class is UST, which faced significant challenges in May 2022 (see Section~\ref{sec:protocolrisk} for more discussion). Unlike traditional stablecoins with fiat reserves, UST depends on algorithms, market incentives, and its sister token, Luna, to maintain its peg to the US dollar.  When the demand for UST increases and its price rises above the peg, new UST tokens are minted, and a corresponding amount of Luna tokens are burned. Conversely, if the demand for UST decreases, UST tokens are purchased from the market and burned, while new Luna tokens are minted to restore balance. This mechanism helps maintain the stability of Terra stablecoins while allowing Luna's value to fluctuate.

\section{Decentralized Exchanges} \label{sec:dexes}
Decentralized exchanges (DEXs) allow users to trade cryptocurrencies without the need for a centralized intermediary. Instead, they use automated market makers (AMMs) to pool liquidity from users and determine prices using an algorithm. Each liquidity pool owns two types of tokens, say tokens $A$ and $B$, and supports the exchange of one token against the other. This means that a user may either purchase token $A$ by paying units of token $B$, or vice versa. The quantity of tokens $A$ needed to purchase one token $B$ and vice versa is determined by the so-called AMM pricing algorithm.

An AMM pricing algorithm implements a function that maintains the liquidity invariant after the occurrence of a trade. Concretely, suppose $x_A$ and $x_B$ are, respectively, the quantities of tokens $A$ and $B$ in the pool. If an investor trades in $\delta_A$ token $A$, the quantities  $\delta_B$ of token $B$ extracted from the pool satisfies the relationship:
$$
F(x_A, x_B) = F(x_A + \delta_A, x_B - \delta_B),
$$
where $F:\mathbb{R}_+^2 \rightarrow \mathbb{R}$ is the pricing function. The marginal exchange rate between tokens $A$ and $B$ is then given by $\frac{F{x_A}}{F_{x_B}}$, where $F_{x_A}$ and $F_{x_B}$ are, respectively, the partial derivatives of $F$ with respect to $x_A$ and $x_B$.

The function $F$ is typically chosen to be a twice differentiable convex function. \citet{CapponiJia} show that the convexity of the function is directly related to the price impact, with a higher convexity capturing a larger impact of the trade on the price.

\begin{figure}[t!]
    \centering \includegraphics[scale=0.75]{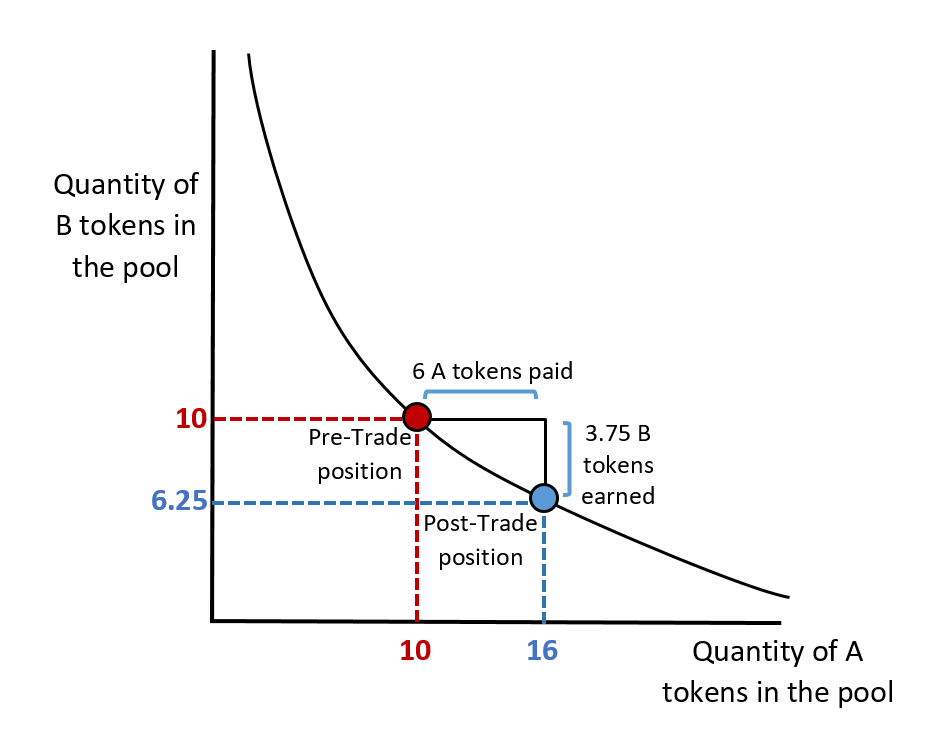}%{pricingcurveNFA2.png}
    \caption{The graph plots the constant product Function, i.e., the pricing algorithm implemented by Uniswap, the decentralized exchange with the largest trading volume.}
    %\vspace{-10cm}
    \label{fig:pool}
\end{figure}

Consider, for example, Uniswap V2 and Pancake Swap, two of the largest DEX protocols which, as of March 7, 2023, account for above 10\% of market share in terms of traded volume.\footnote{See \url{https://www.coingecko.com/en/exchanges/decentralized} for more details.} Both protocols implement the constant product function   $F(x_A,x_B)=x_A x_B$. If $x_A=x_B=10$, and an investor trades in $\delta_A=6$ token $A$, the quantity $\delta_B$ of tokens $B$ she takes out of the pool satisfies the relation $10 \times 10 = 16 \times (10-\delta_B)$. This yields $\delta_B = 3.75$, i.e., the investor does not exchange tokens at the pre-trade marginal exchange rate $\frac{x_B}{x_A} = 1$ but rather at the exchange rate $\frac{x_B-\delta_B}{x_A + \delta_A} = 39\%$. The difference between pre-trade and post-trade exchange rates captures the cost of trading, which in turn depends on the convexity of the constant product function. We graphically illustrate this pricing mechanism in Figure~\ref{fig:pool}. We also refer to \citet{Uniswapwhite} for additional details on the Uniswap AMM.

 %\vspace{5cm}

AMM protocols differ in terms of value locked (also referred to as the liquidity level) and pricing functions used. We refer to \citet{xu2021sok} for a comparison of the four major AMM protocols according to their market share, including Uniswap (both V2 and V3), Balancer, Curve, and DODO. The analysis of AMMs has been subject of considerable investigation recently (see, among others, the works by  \citet{MoallemiRough}, \citet{LeharParlour}, and \citet{HasbrouckAMM}). 

\paragraph{Decentralized vs Centralized Exchanges.} Decentralized exchanges (DEXs) and centralized crypto exchanges (CEXs) operate under fundamentally different mechanisms, with differences in liquidity provision, price setting and updating, transaction matching and settlement, and underlying infrastructure.

In terms of the liquidity provision mechanism, CEXs typically operate a central limit order book (LOB). Highly skilled market makers compete with each other by submitting limit orders, i.e., quantities and prices at which they are willing to buy or sell. These market makers are typically high-frequency trading firms that heavily engage in an arms race on technology, invest in microwave cables, and co-locate their servers with the exchanges in order to exploit arbitrage opportunities at the expense of investors. As argued in \citet{budish}, this arms race for speed is socially wasteful, and the cost are passed to investors in the form of higher bid-ask spreads. By contrast, in a  DEX, any agent regardless of his level of skills and expertise, can supply liquidity to an AMM. Liquidity providers share the fees earned from trading activities and the risk of liquidity provision (such as the risk of being arbitraged, as discussed in \citet{CapponiJia}), without any need of engaging in a  wasteful arms race for speed.

Prices in decentralized exchanges are set and updated differently compared to centralized exchanges. AMMs deploy deterministic pricing functions to set the price schedule. Unlike centralized limit order books where market makers submit limit orders to update prices, prices can only change in response to trades in decentralized exchanges. This is because the token exchange rate is only a function of the pool's inventory, and specifically of the relative supply of tokens. %As discussed in \citet{CapponiJia}, the rigid nature of AMM price schedules makes it important to design the pricing function efficiently.

Transactions are matched and settled differently in AMM compared to CEX. In DEXs, users do not need to be paired to complete a transaction, but gain immediate access to available liquidity through interaction with smart contracts. Moreover, as the transaction is confirmed by validators on the underlying blockchain, it is automatically settled and the corresponding token balances of each transacting party are updated. In contrast, settlement and clearing are typically much more complicated for traditional centralized exchanges. For instance, it takes two business days after the execution date for a US equity transaction to settle.

%CEXs are typically built upon centralized servers, whereas DEXs typically operate on public blockchains. The different supporting infrastructure has the following implications for tradings risks and price discovery. 

In a LOB a market order, i.e., an order to buy or sell a security at the best available price, is executed immediately once placed. Hence and unlike DEXs, market orders are not publicly exposed before execution. Hence, they are not exposed to the risk of being frontrun while pending, which is instead the case for DEXs (see \citet{frontrunning}). 

CEXs support continuous, serial execution of orders. \citet{HasbruckRFS} show that a timestamp resolution higher than one second is needed to analyze  price discovery in the U.S. stock market, because trading is very fast. In contrast, DEXs execute orders discretely in batches as dictated by the underlying blockchain infrastructure. For example, on the proof-of-stake (PoS) Ethereum blockchain, pending orders are executed by validators every 12 seconds. Hence, price discovery on DEXs occurs at fixed time intervals and can have delays.

In a LOB market, limit orders submitted by market makers are aggregated and incoming market orders are executed against the resting limit orders in the book. Hence, LOB prices can adjust as market makers revise their quotes responding to new information such as public news and the process involves no trade executions. For instance, \citet{BrogaardJF} show that price discovery occurs predominantly through limit orders in the Canadian stock market. Unlike LOB markets, liquidity providers do not submit price quotes on AMMs. Instead, they provide liquidity by depositing tokens in the pool and incoming orders are executed against a pre-determined pricing curve such as a constant product function. As a result, AMMs prices can only adjust through trade executions.

Because DEXs execute orders in batches, they require traders to bid a priority fee (e.g., gas fee on the Ethereum network) to determine the execution priority of their orders. Priority fee bidding is a unique feature of trading on DEXs, which can affect the trading behavior of informed traders and thus the price discovery process on DEXs. The study of \citet{CapponiJiaYu} demonstrates that the priority fee reveals private information of traders. They estimate a structural vector auto-regression model to trade data from the largest centralized and decentralized exchanges, respectively Binance and Uniswap, and orders which are pending on the public mempool of the Ethereum
blockchain. They find that DEX trade flows with a high priority fee has a much larger permanent price impact than low-fee trade flows.

Recent research, including studies by \citet{AoyagiIto} and \citet{AMMLOBMoall}, has contrasted decentralized and centralized exchanges. \citet{AoyagiIto}'s study introduces a theoretical model to analyze co-existence of exchanges with diverse market-making structures, namely a centralized limit-order book and a decentralized automated market maker. On the other hand, \citet{AMMLOBMoall} presents a comprehensive framework for designing and analyzing asset exchange mechanisms. This framework unifies and facilitates comparison between the two prevailing paradigms for exchange: AMMs with a constant product function and limit order books.

%Their calibration shows that the permanent price impact of the high-
%fee DEX trade flow ranges between 4.27 and 8.16 basis points, while for the low-fee DEX trade
%flow it is between 0.41 and 0.94 basis points. These estimates remain robust if we control for CEX
%trade flow and size of DEX trades, meaning that the priority fee attached to DEX trades reveals
%private information beyond what is captured by these two other confounding factors.

\section{Decentralized Lending Platforms}\label{sec:lending}

In traditional finance, the bank acts as an intermediary between the demand and supply side of credit. Savers deposit money to the bank and they are promised interest on the deposited funds. The bank lends out funds to firms and households. The critical function of the bank is to assess the credit quality of borrowers. Typically, these steps involve checking borrowers’ credit scores through their borrowing and payment histories, total debt currently owed, and personal information such as income. Some borrowers may be difficult to screen because of the limited information available on their behalf, in which case the bank would require collateral to secure the loan.%, so as to align incentives of borrowers and lenders. 

DeFi lending platforms are marketplaces where depositors and borrowers interact directly with the smart contract, without any need for a bank intermediary. Lenders provide liquidity by depositing tokens into a pool. Simultaneously, borrowers take loans from the pooled funds after pledging collateral. Borrowers and lenders do not need to be individually matched. Rather, loans can be accessed instantaneously based on the requested amounts and posted collateral. 

DeFi loans are typically disbursed in stablecoins, i.e., cryptocurrencies whose value is pegged to another asset class such as fiat currency or gold. The posted collateral consists of riskier crypto assets (see \citet{IMF20022}). Smart contracts assign each collateral type a margin, that determines the minimum collateral borrowers must pledge to receive a loan of a given amount. To deal with the high price volatility of crypto assets and the pseudoanonimity of the trading parties, the loan is typically overcollateralized, i.e., the demanded collateral is higher than the loan size. Collateralization rates typically range between 0 and $80\%$ on major lending platforms. For example, if the collateralization rate is $80\%$, borrowers can borrow up to 80 percent of the collateral value posted. Consider, for example, the decentralized stablecoin DAI. The minimum collateralization ratio for Ether, the native cryptocurrency of the Ethereum ecosystem is currently set at 66\%, i.e., depositing \$150 worth of Ether allows one to borrow up to 100 DAI. If the collateral factor is zero, as in the case of Tether (USDT) in some DeFi platforms, the user cannot borrow using the asset as collateral. 

The DeFi lending platform also sets a “liquidation ratio” relative to the borrowed amount. For instance, an 80\% collateralization rate may be accompanied by a 90\% liquidation ratio. If collateral depreciates and its value falls below this threshold, the contract stipulates that any user of the platform can liquidate and seize the collateral, use the liquidation proceeds to repay the lender and pocket the residual value. %DeFi lending platforms typically charge fees paid by borrowers for their services.

Next, we describe the implementation of Aave, a decentralized finance (DeFi) lending platform operating on the Ethereum blockchain. Aave ranks among the top five DeFi protocols, and as of March 1, 2022 it was accounting for about 13\% of the total value locked in DeFi.\footnote{We refer to \url{https://phemex.com/academy/aave-defi} for additional details.}
The Aave smart contract decides both the interest rate paid by the borrower and the rate received by the lender. The interest rate charged to borrowers depends on the supply of funds available in the pool at a specific time. If the supply of funds decrease, then the algorithm raises the interest rate. The smart contract also employs a reserve factor, i.e., a share of the borrower’s accrued interest that should be deducted and set aside for periods of illiquidity. Hence, the interest earned by lenders equals the interest paid by borrowers minus the reserve factor. We refer to \citet{Aavewhite} for additional details on the lending protocol.

It is noteworthy to highlight the emergence of privacy-preserving oracle protocols in DeFi, which are unlocking opportunities for {\it undercollateralized lending}. These protocols are equipped with a technology which allows retrieving off-chain data, encompassing credit scores, bank account balances, and repayment history, and securely deliver this information  onto the blockchain in a privacy preserving manner. By using this data, smart contracts can then adjust collateral requirements for borrowers with proven creditworthiness. One notable example of such a protocol is DECO (refer to \citet{DECO} for technical specifications), utilizing cryptographic proofs to demonstrate that a borrower's credit score, as evaluated by an established credit bureau, surpasses a specified threshold while safeguarding the actual credit score from disclosure. 

The introduction of undercollateralized lending in DeFi holds the potential to fuel market growth by enhancing capital efficiency in contrast to overcollateralized lending. This shift empowers consumers to swiftly obtain loans from decentralized applications in a matter of minutes, circumventing the frictions of borrowing capital from centralized intermediaries. 

\subsection{Decentralized vs Centralized Lending.} 

Decentralized lending present both costs and benefits to users. One of the main costs in decentralized lending is the need for over-collateralization originated from the pseudo-anonymous nature of the blockchain, which makes it difficult to assess the creditworthiness of borrowers. %Hence, under-collateralized risky lending based on the credit worthiness of the borrow is not implementation.
%To ensure that borrowers do not engage in excessively risky activities leading to their inability of repay, it becomes necessary to over-collateralize the loan. 
Besides being capital intensive, the over-collateralization can also lead to severe downward price pressures on cryptocurrencies if collateral is liquidated during periods of stress, as demonstrated during the Black Thursday. On Thursday, March 12 2020, MakerDAO experienced a total of 3994 liquidations, with over \$10m worth of Ether collateral liquidated on that day alone. This led to a crash of the Ether cryptocurrency, which lost up to 50\% in a single day.\footnote{See \url{https://medium.com/defi-saver/liquidations-in-defi-how-they-happen-and-how-to-prevent-them-9caddd52de71} for additional details on MakerDAO liquidations.} \citet{ParlourLeharsyst} compile a dataset of collateral liquidations on Compound and Aave, the two largest DeFi lending protocols. They find that deleveraging leads to both a temporary and permanent impact on prices. They show that there is a negative feedback loop between liquidation and prices: the liquidation of loans leads to downward price pressure on collateral and, as a result, a higher amount of loans is then liquidated.

Despite the risks, decentralized lending offers several benefits over traditional centralized lending, including transparency, guaranteed enforcement, and lower costs. First, the rules which specified lending and borrowing terms are hard-coded in the smart contract and fully auditable. These rules, including the condition for loan origination (e.g., maximum collateralization rate), the algorithm for setting interest rates, and the trigger for collateral liquidation, are automatically enforced by smart contracts. The automated nature of decentralized lending not only implies very little exposure to counterparty risk for both lenders and borrowers, but also mitigates the moral hazard problem arising from the excessive risk-taking behavior of intermediaries because there is no discretionary decision-making by intermediaries. Assets deposited in the lending pool, originated loans, and received interest payments are publicly observable in real-time. This unprecedented level of transparency has the potential to reduce monitoring costs, because it eliminates the need of human supervision and thus has the potential of simplifying the regulation process. %This reduction in opacity can potentially help prevent systemic crisis such as that experienced in the 2008 financial crisis where assets under management were not transparently observed.
Moreover, the organization, servicing, and and closing of a loan are automatically executed by the smart contract, which not only reduces enforcement costs but also eliminates the need for relationships and manual work in the process.

Decentralized lending platforms also enable a set of services  such as flash loans, which would not be possible to offer on traditional centralized lending platforms. Flash loans are uncollateralized loans without borrowing limits, which allow a user to borrows funds and return them in the same transaction. This service provides potentially infinite leverage to users, and has been used to fund arbitrage strategies, such as arbitrage between on-chain and off-chain exchange rates, that improve market efficiency.

% The mechanical nature of smart contracts also helps coordination in stressed times and eliminates strategic precautionary liquidity hoarding, which typically leads to liquidity freezes. During periods of financial distress, traditional lenders who fear to be lacking funds for their own projects due to a high likelihood of shock occurrence or market tightness, refuse to lend and exacerbate financial distress (\citet{Brunnermierdecipher}). By contrast, the rule governing lending behavior are specified ex-ante in a smart contract, and the distributed nature of the contract verification does not make it possible to unilaterally stop its execution, unless precise conditions for stopping the smart contract were specified in the program ex-ante (see \citet{MakaroShoar} for an in-depth discussion of execution and enforcement in smart contracts).  

\section{Decentralized Governance in DeFi}\label{sec:decgov}

Governance defines the set of rules and procedures that regulates the behavior of all participants in the ecosystem. 
Decentralized autonomous organizations (DAOs) are the most commonly used governance structure for DApps and DeFi projects. DAOs utilize \textit{governance tokens} to distribute power among community members, granting them permission to vote on management and decision-making of their organization.

\subsection{Examples of DeFi Governance}
MakerDAO was one of the earliest projects to issue a governance token (see \citet{Makerwhite}). MakerDAO is a DAO that develops technology for borrowing, lending, and savings on the Ethereum blockchain, and launched the DAI digital currency, a stablecoin pegged to the U.S. dollar and collateralized by crypto assets. The Maker Protocol is governed by holders of its MKR governance token. Each MKR token entitles its holder to one vote. MKR token holders can vote on appointing team members, system parameter changes, technical improvements to the protocol, and operational spending for the system. The MKR voting system follows a proportional rule, meaning a voter with more MKR tokens locked in the contract has a greater decision-making power.

Another example of governance token is the UNI token. This is the native token of Uniswap, and it is used to vote on governance proposals of the Uniswap protocol. Users who hold $1\%$ or more of the total UNI token supply can submit development proposals, such as adding new features to the automated market making protocol or changing the fee structure. Any UNI holder, regardless of how much they hold, is able to vote on such proposals. %One of the proposals has been to grant some utility to holders of the governance tokens. 
As it stands, the entire 0.3\$ fee per traded volume is passed to liquidity providers. A fee switch proposal\footnote{See \url{https://gov.uniswap.org/t/fee-switch-design-space-next-steps/17132} for more details.} submitted in July 2022, also referred to as the protocol charge, is to give $0.25\%$ to liquidity providers and the remaining $0.05\%$ to UNI token holders as a reward for holding the token.

\subsection{Promises and Pitfalls of DeFi Governance}

DeFi governance leverages blockchain technology to enhance the transparency, accuracy, and efficiency of voting. Compared to traditional shareholder voting mechanisms used in publicly traded companies such as proxy voting, DeFi governance is more streamlined and less prone to errors.

 In the United States, most investors hold their stocks through their brokers, who are responsible for tracking ownership and interests, delivering voting material, collecting voting decisions, and casting votes on behalf of their clients. This complex process often leads to technical glitches, such as over-voting, where a broker casts more votes than it or its clients are allowed to. This can occur due to brokers' inaccurate tracking of beneficial owners' interests, such as failure to match shares held in a margin account that the broker lends out, with the corresponding client's diluted voting rights during the re-hypothecation period.

The transparency of ownership tracking in blockchain and the automation provided by smart contracts can solve these imperfections. All users share the same distributed ledger, and the transaction flow history is transparently observed on the blockchain, enabling real-time tracking of voting power. Users who have voting power can directly submit their votes, without intermediaries, which eliminates technical problems due to delayed recording or information transmission, as well as inaccurate recording. The streamlined voting procedure is entirely automated on DeFi, which saves the operational costs that exist in proxy voting.

DeFi governance also allows to implement sophisticated voting structures at reduced costs. In the traditional financial system, a company's board can issue different types of shares to attribute voting rights, such as ordinary shares with one vote per share and executive shares with 100 votes per share. However, these structures are expensive to implement and maintain, mostly because of the high auditing and compliance costs. Governance tokens can be programmed to make voting rights a pre-specified function of token holdings, and smart contracts would automatically enforce the voting rules, thus eliminating the need of costly legal and administrative fees.

As for traditional corporate governance, DeFi governance suffers from the misalignment of interests between governance token holders. Governors of decentralized lending platforms, for example, may also be providers of funds to the pool, and thus want to extract high interest. This creates an incentive for them to vote on proposals that set interest rates way too high to benefit the whole ecosystem. Conversely, governors who want to extract cheap funding may vote for proposals that lower interest rates. Governors who are part of the development team may want to pass proposals that do not bring new talent to the operating team, so that they can remain on board even with a reduced skill set.

Despite the original intention of creating a transparent decentralized governance process, many DeFi projects have evolved towards an oligopolistic  structure where a few accounts holding the vast majority of tokens. This centralization may lead to collusion, manipulation, and embezzlement by insiders. We refer to \citet{walch} for empirical evidence of these forms of unethical behavior. For example, Wonderland, a decentralized autonomous organization (DAO) that raises capital to invest in DeFi, suffered from embezzlement by its former Treasury manager, Sifu. Sifu, who accumulated many governance tokens for Wonderland, later founded Sifu's Vision project, which issued Sifu tokens with no intrinsic value. Sifu was the second biggest voter, and his accumulated governance tokens allowed Wonderland to pass a vote to buy \$25 million worth of Sifu Vision tokens, resulting in a theft of \$25 million from Wonderland DAO.\footnote{We refer to \url{
https://cryptobriefing.com/sifu-forces-wonderland-to-give-him-25-million/} for additional details.} 

Smart contracts provide a powerful tool for mitigating centralization risks in DeFi governance. One approach is to design voting structures that limit the power of large token holders. For example, one might conceive a voting structure where the number of allowed votes is a concave function of the number of held tokens, i.e., the marginal voting power of a token holder decreases as the number of his tokens increase. This governance structure would help prevent a small group of large token holders from dominating the decision-making process. However, this also gives big voters incentives to split their tokens across different accounts. An alternative approach is to reward loyalty by tracking the amount of time a governance token has been held before the voting decision. This would help prevent strategic behavior of users who purchase governance tokens solely for the purpose of voting on a particular proposal. In addition, hard-coded quorums can be built into the voting rules, requiring a minimum number or percentage of voters to participate in the vote. This would ensures that voting power is not excessively concentrated, but also that a sufficient level of community engagement is attained.

\subsection{Mitigating CeFi Governance Risks through Decentralized Technologies}

Centralized crypto firms are often not required to disclose audited financial statements, and they may intentionally avoid regulations that publicly traded companies are subject to regarding internal control, corporate governance, and financial disclosures. This lack of transparency and accountability can lead to corporate mismanagement, which puts users' funds at risk. A noticeable example is the recent FTX’s chapter 11 filing which, according to the appointed CEO John Ray III in charge of the FTX’s governance structure, was deemed to be the worst in his 40 years of restructuring experience. John Ray noted various instances of mismanagement of cash and inadequate financial reporting, including granting privileged trading terms to Alameda, an affiliated proprietary trading firm, such as relief from automatic liquidations of collateral when margin calls were made.

Defi offers tools to mitigate corporate mismanagement, such as the misuse of users' funds. These problems can be mitigated through the decentralized custody of assets, which can be implemented through multiparty computation (MPC). The private key of a digital wallet is then shared among multiple parties, with each party in possession of a portion of the key acting as an approver. This mitigates the risk that an internal bad actor can gain access to a full key and misuse assets. MPC protocols were first introduced by \citet{Yao}, who considered the two-party case, and then generalized to the multi-party case by \citet{Micali}. They combine cryptography with secure communication protocols to allow different parties to jointly manage users’ digital assets. The process involves complex calculations and communication between the parties, which can be computationally intensive and consume a lot of results, but it results a secure and efficient way to manage users’ cryptocurrency assets.

\chapter{Operational Risks in DeFi}\label{sec:operationalrisk}

Operational risk refers to the potential for failure coming from manipulation, or breakdowns in internal procedures and systems due to external events. In DeFi, this risk can be broadly classified into five main categories: consensus mechanisms risks, protocol risk, oracle risk, frontrunning risk, and systemic risk.  

\section{Consensus Mechanism Risks}\label{sec:consensusrisk}
%Consensus mechanism risks refer to vulnerabilities of consensus protocols used in blockchain, such as Proof of Work or Proof of Stake. 

%Protocol risks are unique to the DeFi ecosystem, and can be broadly classified into two categories: risks from the blockchain consensus protocols and risks from the DeFi application protocol.
The reliability of DeFi services hinges on the security and stability of the underlying blockchain consensus mechanisms. Two of the most prevalent consensus mechanisms in DeFi blockchains are Proof of Stake (PoS) and Delegated Proof of Stake (DPoS).

In a Proof of Stake mechanism, as exemplified by Ethereum, an algorithm selects validators for block creation based on the amount of tokens that holders stake from their crypto-asset ownership. The process involves three steps: selecting a proposer, proposing a block, and then validating the proposed block. Validators with higher staked amounts have a greater likelihood of being chosen by the algorithm (see \citet{SalehPoS} for an analysis of PoS algorithms). However, this direct correlation between ownership and selection chances can create centralization risks within PoS consensus mechanisms.

To address some of the centralization concerns, Delegated Proof of Stake (DPoS) introduces a democratic element to the PoS consensus mechanism by outsourcing the validation process. Similar to PoS, block validation is still randomized, and individuals or entities (stakeholders) with the highest stake of a particular crypto asset are more likely to be selected for block validation and rewarded. The crucial difference lies in the DPoS model's implementation of a voting system. Validators chosen to validate a block can delegate their work to third-party entities responsible for achieving consensus during block generation and validation. This delegation promotes greater decentralization. However, a potential issue arises when stakeholders with larger stakes collude to form cartels, leading to increased centralization in the network.

Centralization gives an undue amount of control to a few entities, allowing them to wield significant influence over the network's decision-making processes. As a network becomes more centralized, it becomes increasingly vulnerable to attacks, such as double spending. In this attack, counterfeiters spend a cryptocurrency in one block and later undo this expenditure by releasing a forged blockchain in which the transaction is erased. Such an attack is usually carried out through a fork, in which the attacker creates competing branches, each registering a potentially different version of the ledger. Fork attacks can affect the finality of payment or DeFi transactions, which can no longer be considered immutable. With the increasing adoption of DeFi, incentives for the execution of fork attacks have become even stronger. These attacks are generally associated with transaction fee-based forks and time-bandit attacks (\citet{frontrunning}), which lead miners to earn additional income from optimizing the initial ordering of transactions in a block and rewriting the history.\footnote{After the switch to proof of stake consensus protocols, Ethereum imposed a punishment on validators who engage in unethical behavior. This process is referred to as slashing, and results in forcefully removing the dishonest validator from the network with an associated loss of all their staked Ether tokens.}

%https://www.sciencedirect.com/science/article/pii/S1319157822003792#b0230

%https://www.bis.org/publ/work765.pdf and file:///C:/Users/ac3827/Downloads/SSRN-id3108601%20(1).pdf
\section{Protocol  Risks}\label{sec:protocolrisk}

DeFi protocols use smart contracts to replicate traditional financial services on decentralized ledgers, and the protocols of these applications also have risks.  Coding vulnerabilities, such as arithmetic errors, casting errors, inconsistent access control, and coding bugs in smart contracts, are often exploited by attackers, and they aim to alter the protocol's behavior.  A prominent example is the DAO, a decentralized autonomous organization launched in April 2016 via one of the largest crowdfunding campaigns of token sales in the history. The DAO's protocol was hacked three months after its launch and \$60 million of Ethereum was stolen. The Ethereum blockchain was eventually hard forked, i.e., the entire history of transactions was written, and this allowed restoring the stolen funds. \citet{DeFiattacks} compile a dataset consisting of 117 and 69 incidents on the Ethereum and Binance smart chain, respectively. They  show that hacks resulted in a total loss of at least 3.24 billion USD from Apr 30, 2018 to Apr 30, 2022 in the DeFi ecosystem. They find that the majority of the DeFi incidents occurred after late 2020, with the peak being reached in August 2021 when nearly 600 million dollars were lost.  

The case of DAO indicates that forks can be used to bring the system back to a reliable state and prevent misappropriation of funds.  Nevertheless, forks affect the finality of DeFi transactions and thus undermine of the foundational principles of blockchain, namely the immutable nature of the ledger where contract agreements, once settled, are also final.

Other types of attacks on DeFi protocols resemble financial market manipulation, which often exploits market design flaws, or unsafe external protocol dependency or interactions with other DeFi protocol. A notable incident is the crash of the Terra-Luna ecosystem, which provided UST, a decentralized, algorithmic  stablecoin pegged to fiat currencies; see also Section~\ref{sec:stablecoins} for more details on the UST stable coin. The Terra-Luna system relies on the incentives of arbitrageurs in exploiting temporary relative price discrepancies between Luna and UST tokens. Investors can exchange 1 UST for \$1's worth of Luna at any time. If the price UST exceeds its peg, arbitrageurs can buy Luna, swap it for UST at the current market price, and then sell the UST for the pegged fiat currency, earning a profit. This process increases UST supply and decreases Luna supply, driving UST's price back towards its peg. Conversely, when the price of UST falls below its peg, arbitrageurs can buy UST at a discount, swap it for Luna, and then sell Luna for a profit. 

However, a key design flaw  in the system is to assume that minting more of an asset would lead to a proportional increase in its total market capitalization. In May 2022, a large UST sell-off caused its price to drop to \$0.91. In response, the system minted a significant number of Luna tokens to stabilize the price and implemented multiple measures to counteract the decline\footnote{See \url{https://www.coindesk.com/markets/2022/05/10/usts-bitcoin-reserve-too-late-in-coming-to-save-dollar-peg/} for additional context.}. However, when Luna's price fell below a specific threshold and UST's price failed to revert to its peg, a "run" on both UST and Luna ensued as more people and institutions began to liquidate their holdings. This led to a further decline in the prices of UST and Luna until both tokens became virtually worthless. We also refer to \citet{MakarovShoar}, and Appendices C and D of \citet{Uhlig} for additional details on the Terra blockchain ecosystem and its crash.

\section{Oracle Risk}

Many DeFi protocols rely on external information as input. Such information is provided by oracles, namely applications that source, verify, and transmit external data to smart contracts running on the blockchain.

Oracles are a critical component of smart contract execution. One of their main uses is feeding information to stablecoin protocols. For example, the Maker Platform requires real-time information about the market price of assets used as collateral against borrowing of the decentralized stablecoin DAI. Such information is necessary to decide when to trigger liquidations. However, stablecoins built using smart contract technology typically have no direct access to price feed data regarding their exchange rate to USD or the values of posted collateral. Rather, they rely on oracles to provide such information.  
%on-chain collateralized and algorithmic stable coins
There are two types of oracles: centralized and decentralized. Centralized oracles act as a single trusted entity that provides data from an external source to a smart contract. As such, they are vulnerable to single points of failure or manipulation, because a single entity controls all updates. Decentralized oracles, on the other hand, rely on multiple sources of data to achieve consensus, without a single trusted party. They may potentially be less prone to manipulation, but it is not immune to risk.

Numerous solutions are available for implementing decentralized oracles. One approach involves leveraging consensus mechanisms, where information is sourced from multiple external providers, and a consensus is reached to ensure accuracy and trustworthiness.

Providers are rewarded for providing valid information. One such example is Chainlink (see also \citet{Chainlink} for their whitepaper), which facilitates the transfer of data from off-chain sources to on-chain smart contracts. Ideally, all sources report data without coordinating with one another, but most oracles relying on multiple parties are also vulnerable to collusion between reporting parties. Some other oracles feed information directly to other on-chain DeFi applications. A prominent example of this type of decentralized oracle is Uniswap, which directly feeds on-chain price information and liquidity data of DEXs into other DeFi protocols such as Aave or Compound, which implement borrowing and lending. In this way, price information is determined jointly by liquidity providers and traders on DEXs, but is not controlled by a single party. One potential risk for this type of oracles is that, if the DEX application used to feed information is manipulated or attacked, then information it feeds will also be tampered with.

Both decentralized and centralized oracles are vulnerable to attacks, and these have caused massive losses to the DeFi ecosystem. (see also \cite{DeFiattacks})\footnote{A recent example of a costly oracle attack is discussed in the article ``DeFi Lending Protocol Fortress Loses All Funds in Oracle Price Manipulation Attack.'' from \url{https://cryptonews.com/news/
defi-lending-protocol-fortress-loses-all-funds-oracle-price-manipulation-attack.htm}.} \citet{qin2020attacking} provide a detailed analysis of two popular oracle manipulation attacks, one of which was a decentralized oracle. The attacker borrowed a flash loan
of 7,500.00 ETH, and converted a total of 4,417.86 ETH to 1,099, 841.39 synthetic USD tokens, also referred to as sUSD. By doing so, he distorted the exchange rate on Uniswap and Kyber from 268.30 sUSD/ETH to 108.44 sUSD/ETH, while other DeFi platforms remained unaffected at 268.30 sUSD/ETH. Uniswap and Kyber were the two DeFi services to provide price information to the lending platform where the flash loan was borrowed. The attacker then
collateralized all the 1, 099, 841.39 purchased sUSD to borrow 6, 799.27 ETH at an exchange rate of 162.66 sUSD/ETH on such lending platform. After this transaction, the adversary possessed 6,799.27 plus the 3,082.14 ETH which were not converted to sUSD tokens. He then repaid the flash loan amounting to 7,500.00 ETH, and was able to generate a revenue of 2,381.41 ETH while only paying 0.42 ETH transaction fees.

To mitigate risks, decentralized oracles have to be carefully designed. One strategy involves utilizing statistics that are inherently resistant to manipulation. An example of such a statistic is the Time-Weighted Average Price (TWAP) offered by Uniswap's oracle. Manipulating TWAP poses significant challenges as it would require alteration of price information recorded in previously confirmed blocks. This can only be achieved by forking the original chain, an operation that carries considerable costs.

Decentralized oracles which rely on multiple sources of information can, for instance, take the median of the reports submitted by all nodes. This statistic is less sensitive to outliers compared to the empirical mean. However, a recent study of \citet{Leifu} argues that the median algorithm may fail to achieve consensus even without an attacker, and ignores the intrinsically high-dimensional structure of decentralized oracles, where each node usually covers many cryptocurrencies. Another approach is to design  schemes which make incentive compatible for reporters to report truthfully. For instance, reporters could stake valuable assets in a smart contract and be punished if they report the wrong data, and rewarded if they report the truth. More research is needed to design incentive-compatible schemes that minimize price manipulation and collusive behavior.

\section{Frontrunning Risk}

In a public blockchain, both in-transit orders and executed trades are transparently observed.  After being submitted and broadcast through the network, orders reside in the public mempool of each validator. While in-transit and waiting to be selected for execution, orders are observable by all participants of the ecosystem. Once it is executed, the order is inserted into a block of the ledger and thus becomes part of the recorded history of transactions.

The transparency makes it possible to trace and verify transactions in real time, and thus enables parties to fully trust each other without any need of an intermediary. However, malicious agents can capitalize on the information content of the order, and extract value  through a frontrunning attack to the user who submitted it. Losses generated from frontrunning transactions have been very high (see \citet{eskandari2019sok}), and the risk of frontrunning is widely recognized as one the main obstacles behind the broad adoption of DeFi technology.

Frontrunning attacks include displacement, sandwich, and suppression (see \citet{torres2021frontrunner} for a detailed treatment). The most common type of attack in today's DeFi ecosystem is the sandwich attack, which works as follows.\footnote{In traditional finance, frontrunning, i.e., the practice where traders or brokers execute a trade before a prior large order is executed, is highly illegal and unethical.} Suppose an attacker observes a pending order to exchange Ether for USDC tokens in a decentralized exchange. She will then broadcasts one frontrunning and one backrunning transaction to sandwich this order. The frontrunning transaction would exchange a large quantity of Ether for USDC, and bid a higher fee than the victim transaction. The backrunning transaction would exchange USDC for Ether, and bid a lower fee than the victim transaction. Clearly, the transaction with a higher fee will be prioritized by the validator, and thus the attacker will buy USDC at the Ether price which should have been instead paid by the victim user. The execution of this transaction will lead to a price increase of USDC relative to Ether. Hence, the victim will receive a lower quantity of USDC tokens for the deposited Eth tokens. After the victim's transaction has been executed, the frontrunner will then make revenues from executing her reverse transaction. 

\citet{ParkAMM} highlights several operational characteristics that introduce conceptual complexities to automated market making. These include the absence of time priority for trades, the capacity for third parties to observe unsettled trades, the uncertain price determination prior to settlement, and the inherent profitability of front-running given all recognized hard-coded pricing functions. A study by \citet{CappMEV} shows that frontrunning risk is a friction, which prevents from attaining an allocation of blockspace to transactions which maximizes the value o the ecosystem.  %, and increases congestion in the blockchain netwo
%rk. 
Users have weaker incentives to submit frontrunnable transactions, which reduce the aggregate welfare of the ecosystem if these  transactions have a high value for the submitting users. Moreover, 
%The second inefficiency originates from the fact that frontrunning transactions increase congestion in the network, as the number of submitted orders goes up. 
a frontrunning order would only result in a transfer of wealth from the victim user to the attacker. Hence, allocating block space to such an order would only result in a waste of resources without generating value for the system. %\citet{CappMEV} design a theoretical model to capture  are supported by empirical evidence, according to which above 30\% of existing blocks allocate space to frontrunning transactions.

Attackers engage in a fierce competition to frontrun an order, because they can appropriate value  only if they are the first to execute. In a competitive market, they are willing to bid a gas fee as high as the value gained from their frontrunning order. The fee revenues are earned by the validators, and they are also referred to as the miner extractable value (MEV).  According to \citet{qin2021quantifying} (see also \citet{BIS} for a follow-up study), the total MEV has reached an estimated amount of USD 550–650 million on just the Ethereum network. However, this is an upper bound for the MEV from frontrunning risk, because their estimates also include MEV resulting from
from forced liquidations, replay attacks, and arbitrage between CeXs and DeXs as studied in \citet{CapponiJia}. We also refer to \citet{frontrunning} for additional background on frontrunning risk, MEV, and gas fee surges originated from frontrunning attacks.

Technological solutions aimed at mitigating frontrunning risk target the protection of order information content. One way to hide the content from malicious attackers is to create an off-chain centralized relay service, which acts as a private communication channel between validators and users. The relay service receives transactions from users, and forwards them to validators, without broadcasting them on the peer-to-peer network. The relay platform screens validators before they join and monitor their activities to prevent that they engage in any frontrunning behavior. One of the most prominent implementations of private submission channels is Flashbots Protects\footnote{For an overview of the Flashbots relay, refer to \url{https://docs.Flashbots.net/Flashbots-auction/overview/}}. Since the first Flashbots block was first mined on December 29, 2020, there has been a steady increase in the adoption rate, which has exceeded $60\%$ in July, 2021 (see Figure 3 in \citet{CappMEV}). This effectively means that $60\%$ of blocks mined on Ethereum contain transactions submitted to the Flashbots relay service.

If these relays are fully trustworthy, and validators and users fully adopt the relay service, frontrunning would no longer be a concern. However, because the use of off-chain private relay channels cannot be enforced in a decentralized environment, it remains to be determined whether users and validators have strong incentives to adopt. If few validators monitor the private channel, users have weak incentives to submit their transactions through such a channel because of the high execution risk faced. Specifically, the likelihood that the validator appending the next block is on the private channel would be low considering that the majority of them only monitor the public mempool. 

\citet{CappMEV} design a game theortical model to characterize under which conditions validators would find it incentive compatible to adopt. They show that if frontrunning risk is not sufficient high, validators have weak incentives to coordinate on a full adoption because they do not want to forgo  MEV. %i.e., the fee revenue coming from frontrunners who identify frontrunnable opportunities and bid to exploit them. 
However, if frontrunning risk is high enough to discourage users whose transactions are frontrunnable from submitting, all validators would monitor the relay service. Frontrunnable users would  then submit their transactions through the relay service, and validators earn the fee paid by them.

To align the private incentives with the social optimum, \citet{CappMEV} propose a contractual solution where users whose orders can be frontrun make an ex-ante commitment to pay an additional fee to incentivize validators to monitor the private channel and execute their orders. This fee serves to compensate validators for the MEV loss incurred if they join private pools. In such case, no frontrunning is possible and thus the MEV originated from gas fee bids of attackers no longer exists.

While private relay channels are the most widely used technological solutions against frontrunning, there have been other proposed approaches in the literature. \citet{Canidio} propose a commit-reveal protocol which is costly, because it requires sending two messages for each transaction, but which can prevent severe front-running attacks while preserving legitimate competition between users. \citet{Aune} propose a technological solution which resembles a zero-knowledge proof mechanism, and consists in publicly broadcasting only a hash of the transaction, but not the exact content so to make it less likely to be attacked. Another technological solution has been introduced in the decentralized exchange aggregator 1inch. Through the feature RabbitHole, the aggregator checks for transactions that are likely to be subject to sandwich attacks. If it determines that a transaction is likely to be attacked in this way, RabbitHole submits the transaction directly to a validator, such as Flashbots, avoiding it being broadcast publicly. We refer to \citet{Parkes2022} for a detailed overview of technological solutions against frontrunning.

\section{Systemic Risk}

In the DeFi ecosystem, the initiation and propagation mechanisms of systemic risk bear resemblance, yet also exhibit differences when compared to traditional financial markets. 
 As in traditional finance systems, distress at one institution can propagate and lead to failure of others (see \citet{Capponisurvey} for a comprehensive survey). 
 
\begin{figure}[t!]
    \centering \includegraphics[scale=0.38]{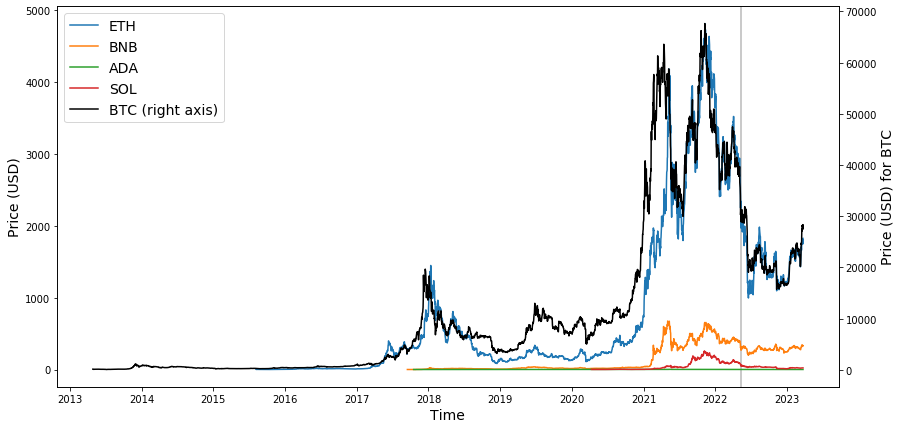}
    \caption{The graph plots the time series of prices of five major crypto tokens: Ether (ETH), Binance Coin (BNB), Cardano's native cryptocurrency (ADA), Solana (SOL), and  Bitcoin (BTC).  We incorporate a dual y-axis which gives a different scale for the USD price of the BTC cryptocurrency, so that the BTC price can be visualized on the same scale as the other crypto tokens.
    As evident from the graph, all prices moved down during the collapse of the Terra ecosystem. We highlight with the distressed period of Terra, ranging from 05/07/2022 till 05/13/2022, with a shadowed area. We source price data of the five crypto tokens from CoinGecko.}
    %\vspace{-10cm}
    \label{fig:pricedrop}
\end{figure}

However, the distinct features of the DeFi ecosystem contribute to the emergence of new systemic risk triggers. Among these is protocol failure, serving as a uniquely inherent source of systemic risk within the DeFi framework.  As discussed in earlier sections, the bad market design of the Terra stablecoin protocol led to the collapse of the Terra ecosystem in May 2022. This collapse wiped out billions of dollars worth of assets on the Terra blockchain, and led to a halt of the Terra blockchain. The collapse of Terra led to a loss of confidence in stablecoins and DeFi. Many users withdrew their tokens from DeFi protocols, which in turn tightened liquidity. Because Terra was deeply integrated across the  crypto ecosystem, its failure triggered a series of price drops of other major cryptocurrencies. This pattern is highly visible in Figure~\ref{fig:pricedrop}, where one can clearly see the sharp simultaneous price drop of 
Ether, Binance Coin, Cardano's native cryptocurrency, Solana, and Bitcoin.

Oracles can act as both initiators and propagators of systemic risk. Issues such as hacks, coding errors, and oracle manipulation could serve as initial shocks, triggering a cascade of effects throughout the DeFi ecosystem. In the event that oracles relay incorrect information, or the information update lags excessively, it could lead to DeFi stablecoin or lending protocols generating inaccurate prices. This could subsequently prompt a substantial liquidation of collateral, potentially causing a systemic collapse and inflicting significant losses on ecosystem participants. For example, the wrong price feeds of the Chainlink oracle during the Terra-Luna collapse led to the collapse of many DeFi protocols (see \citet{Akansha_2022}). 
 
Relative to traditional financial markets, they speed up the propagation of systemic risk and eliminate any discretion in the decision making process. In traditional markets, during periods of crisis, the mark-to-market valuation necessitates some time to fully incorporate all market information, particularly in the event of forced asset sales. On the other hand, in DeFi, an oracle would rapidly incorporate mark-to-market value information into protocols, thereby supplying immediate actionable data. This could prompt swift liquidation of assets and collateral in reaction to negative price shocks. 
%Multiple contagion channels are created from the high interconnectivity of DeFi protocols through oracles. %The collapse of one asset or protocol 
%will quickly feed into other DeFi protocols through oracles, leading to a large amount of liquidation in a short period, which leads to further distress. 
For instance, after the de-pegging of the stable coin UST on May 11, 2022, the value of liquidated collateral on the MakerDao protocol surged from approximately \$400,000 on May 10 to \$9.51 million and \$15.28 million on May 11 and 12, respectively. Given that MakerDao relies on oracles for price information, this large increase  underscores how rapidly systemic distress can spread throughout the ecosystem via oracles.\footnote{We refer to  \url{https://maker.blockanalitica.com/liquidations/?tab=per-date} for a reliable historical source of daily collateral liquidation at MakerDao.} As MakerDao relies on oracles to supply price information, this incident demonstrates the speed at which systemic distress can disseminate across the ecosystem via oracles.

The DeFi ecosystem is particularly susceptible to price contagion effects. Most users maintain portfolios focused on a limited set of highly correlated crypto assets, including Ethereum, Bitcoin, and stablecoins such as DAI, USDC, and USDT. When these users liquidate their portfolios in response to an asset price drop, it initiates a chain reaction, triggering liquidation of other assets. This cascade results in spillover effects and subsequent liquidations, further depressing prices.

 \citet{ParlourLeharsyst} provide empirical evidence
 that collateral liquidation on the main decentralized lending platforms - Compound and Aave - impose high-frequency temporary and permanent price impacts on nine different decentralized exchanges. 

\chapter{Concluding Remarks and Future Research Directions}\label{sec:conclusion}
% In this paper, we have provided an overview of the decentralized finance ecosystem. We have described the functioning mechanism of smart contracts, and explained how they enable the implementation of decentralized protocols for exchanges, borrowing, and lending services. We have described the DeFi governance structure, and explained the mechanisms through which governance tokens allow to transfer control of the protocol from a centralized party to a sett of users who actively participate in it. We have discussed how technological risks, such as hacking and manipulation of oracles and protocols, can become systemic and impose costs on the entire ecosystem. Our analysis highlights the need of further research in this space to improve the design and interoperability of existing protocols, and to guide the regulation of smart- contract enabling blockchain. 

This paper offers a comprehensive overview of the decentralized finance ecosystem. We have focused primarily on the mechanics of smart contracts, decentralized governance structures, and the operational risks that pervade the DeFi ecosystem. 

The majority of assets in the present DeFi ecosystem possess on-chain provenance. We posit that future enhancements to the DeFi ecosystem should foster a closer integration between blockchain technologies and real-world assets, thereby creating an efficient and transparent system for asset ownership and transfer. This can be realized through tokenization - a process that digitalizes a real asset into tokens on a blockchain. Digital tokens backed by real assets are then controlled and executed using a smart contract. 
%There are two types of tokenized assets, namely {\it fungible} and {\it nonfungible}. Fungible tokens include those which are interchangable, because each unit of the tokenized asset has the same market price. For instance, all units of the stable coin USDC are digital tokens pegged to the value of one USD. They are all part of the same network, and each unit is exchangeable for any other.
%Nonfungible tokens, typically abbreviated with NFTs, are not interchangable. Each token has a unique value and cannot be replaced by other tokens of the same type. 
Initial endeavors in this direction include digital representations of commercial real estate. Imagine, for instance, that you possess a California property valued at \$1,000,000. You could tokenize this property, converting it into \$1,000,000 tokens on a public ledger. By acquiring tokens through a DeFi marketplace, an investor essentially purchases a stake in the asset. Details pertinent to the asset are stored in a smart contract, encompassing ownership information, the real estate location, investors' rights, and parties eligible for royalties each time the token is sold.
Asset tokenization has to potential to enhance accessibility,  allowing investors to purchase corresponding digital tokens with increased flexibility in terms of investment proportions and duration. %Asset tokenization allows potential investors with flexibility in terms of investment shares and duration.
A prominent example is the pilot project launched by Ejara, the first and most established decentralized investment and savings platform in Francophone Africa, to overcome barriers to access of the banking system in Cameroon (see \citet{Mercy20222}). In Cameroon, individuals are often faced with the challenge of substantial management or entry fees, ranging from \$10,000 to \$50,000. Additionally, some banks with limited digital services charge clients simply to check their savings accounts. The conventional financial system in Cameroon is primarily paper-based, and financial institutions often try to reduce the costs associated with onboarding new clients (known as KYC costs) by focusing on larger accounts. This approach restricts access to financial services for a broad market, which includes smallholder farmers and marginalized rural populations.

To address these issues, Ejara offers fractionalized government bonds in the form of savings plans. Essentially, Ejara purchases government bonds from the Bank of Central African States and tokenizes them, making them accessible for the wider market. These bonds can be purchased in denominations of 1,000 CFA (equivalent to \$1.53 USD), with maturities ranging from 26 weeks to several years. Higher yields are associated with longer maturities. Each bond is managed by its own smart contract, stored and maintained on Ejara's blockchain. Consequently, Ejara's system allows broader market participation at low entry costs and eliminates the need for entry or management fees, as all processes are automated via the DeFi blockchain's smart contract mechanism.

Further research is needed to enhance the structure and cross-functionality of current protocols. Equally important is the development of transaction fee models and block allocation methods that can foster a more effective distribution of block space. As illustrated in \citet{CappMEV},  current solutions like private submission channels do not eradicate MEV or deter frontrunning attacks. Therefore, alternative solutions that redesign the communication protocol among nodes are needed. For instance, employing transaction encryption strategies using versions of the zero-knowledge proof algorithm, and utilizing fair sequencing algorithms to reorder transactions within blocks, could be promising avenues for exploration.

% An important research direction concerns the design of mechanisms which lead to an efficient allocation of block space. As shown in \cite{CappMEV}, existing solutions, such as off-chain private channels between users and miners, are not able to eliminate MEV extraction and disincentivize frontrunning attacks. Alternative solutions may include revisiting the communication mechanism between nodes by  encrypting transactions in the mempool through variants of the zero-knowledge proof algorithm, or to re-order transactions in blocks using fair sequencing algorithms. 

One significant challenge facing the current DeFi ecosystem is the scalability of the underlying blockchains, which, in their present state, lack the capacity to handle transaction volumes comparable to those in traditional finance. As an illustration, the Ethereum blockchain can currently process roughly 200 transactions per block\footnote{To put things in perspective, compare these estimates with the capacity of the digital payment technology company VISA. The VISA network has the capacity to handle 24,000 transactions per second.}, with a new block being proposed approximately every ten seconds. Strategies like sharding are being explored to enhance processing capacity. However, the optimization of smart contract allocation across various shards remains a complex issue. This is aimed at improving transaction speed and throughput while minimizing cross-shard transaction flow and issues related to asynchronous execution.
Solutions like sharding are being considered to increase processing capacity. It remains to be understood how to optimally allocate smart contracts to different shards to increase transaction speed and throughput while minimizing transaction flow across different shards and asynchronous execution problems.\footnote{We refer to \url{https://ethereum.org/en/developers/docs/scaling/} for technical details of sharding in the Ethereum blockchain.}

Finally, automating the auditing process for smart contracts remains a crucial challenge to be addressed. As highlighted in this paper, the potential vulnerabilities within smart contracts can be exploited by malicious entities, which ultimately erodes user trust and confidence in the DeFi ecosystem. A necessary step towards increasing the adoption of DeFi services involves providing users with a safety guarantee that their transactions and assets are safeguarded from hacking attempts. This security assurance demands a thorough formal verification process, encompassing both contract-level and program-level specifications. Contract-level verification should scrutinize the functional designs of the protocol to evaluate its susceptibility to manipulation. On the other hand, program-level verification aims to audit the source code or compiled bytecodes to confirm the accurate implementation of DeFi protocols. We refer to \citet{taxonomyformal} for a survey of existing smart contract formal specifications and verification approaches.

\section{Acknowledgements}
We are grateful to Ruizhe Jia for excellent research assistance.

\backmatter
\printbibliography

%\bibliographystyle{ACM-Reference-Format}
%\bibliography{book}
%\bibliography{booktest}

\end{document}